\documentclass[]{spie}  
\usepackage[]{graphicx}

\title{NIMBUS: The Near-Infrared Multi-Band Ultraprecise Spectroimager for SOFIA} 

\author{Michael W. McElwain\supit{a}, Avi Mandell\supit{a}, Bruce Woodgate\supit{a}, David S. Spiegel\supit{b}, Nikku Madhusudhan\supit{c}, Edward Amatucci\supit{a}, Cullen Blake\supit{d}, Jason Budinoff\supit{a}, Adam Burgasser\supit{e}, Adam Burrows\supit{d}, Mark Clampin\supit{a}, Charlie Conroy\supit{f}, L. Drake Deming\supit{g}, Edward Dunham\supit{h}, Roger Foltz\supit{a}, Qian Gong\supit{a}, Heather Knutson\supit{i}, Theodore Muench\supit{a}, Ruth Murray-Clay\supit{f}, Hume Peabody\supit{a}, Bernard Rauscher\supit{a}, Stephen A. Rinehart\supit{a}, Geronimo Villanueva\supit{j}
\skiplinehalf
\supit{a}NASA Goddard Space Flight Center, Greenbelt, MD, USA; \\
\supit{b}Institute for Advanced Study, Princeton, NJ, USA; \\
\supit{c}Yale University, New Haven, CT, USA; \\
\supit{d}Princeton University, Princeton, NJ, USA; \\
\supit{e}University of California, San Diego, La Jolla, CA, USA; \\
\supit{f}Harvard-Smithsonian Center for Astrophysics, Cambridge, MA, USA; \\
\supit{g}University of Maryland, College Park, MD, USA; \\
\supit{h}Lowell Observatory, Flagstaff, AZ, USA; \\
\supit{i}California Institute of Technology, Pasadena, CA; \\
\supit{j}Catholic University of America, Washington, DC, USA. \\
}

\authorinfo{Send correspondence to M.W.M.\\M.W.M.: E-mail: michael.w.mcelwain@nasa.gov, Telephone: +1 310 871 7877\\}

\begin{document} 
  \maketitle 

\begin{abstract}
We present a new and innovative near-infrared multi-band ultraprecise spectroimager (NIMBUS) for SOFIA.  This design is capable of characterizing a large sample of extrasolar planet atmospheres by measuring elemental and molecular abundances during primary transit and occultation. This wide-field spectroimager would also provide new insights into Trans-Neptunian Objects (TNO), Solar System occultations, brown dwarf atmospheres, carbon chemistry in globular clusters, chemical gradients in nearby galaxies, and galaxy photometric redshifts. NIMBUS would be the premier ultraprecise spectroimager by taking advantage of the SOFIA observatory and state of the art infrared technologies.  

This optical design splits the beam into eight separate spectral bandpasses, centered around key molecular bands from 1 to 4 $\mu$m. Each spectral channel has a wide field of view for simultaneous observations of a reference star that can decorrelate time-variable atmospheric and optical assembly effects, allowing the instrument to achieve ultraprecise calibration for imaging and photometry for a wide variety of astrophysical sources. NIMBUS produces the same data products as a low-resolution integral field spectrograph over a large spectral bandpass, but this design obviates many of the problems that preclude high-precision measurements with traditional slit and integral field spectrographs.  This instrument concept is currently not funded for development.
  
\end{abstract}



\section{INTRODUCTION}
\label{sec:intro}  

Near-infrared (NIR) astronomy is one of the most exciting windows for viewing our Universe, and new technologies are enabling observers to explore this region of the electromagnetic spectrum with unprecedented sensitivity and precision.  The Stratospheric Observatory for Infrared Astronomy (SOFIA) is uniquely able to access regions of the NIR that are nearly opaque from the ground, particularly those obscured by bands of astrophysically important molecules.  We present the Near-Infrared Multi-Band Ultraprecise Spectrophotomer (NIMBUS) instrument concept that would observe within these telluric bands with unprecedented precision.  NIMBUS would provide the ability to characterize the atmospheres of nearby exoplanets and brown dwarfs, to better understand the star formation in our galaxy and to probe the chemical diversity among primitive bodies in the Solar System.  We note this instrument concept is \textit{not} funded for development.

NIMBUS would be a facility-class NIR (1.34-3.39 $\mu$m) multi-band wide field imager for SOFIA, providing simultaneous high-precision photometry of multiple wavelength bands over a wide field of view (FOV).  NIMBUS's eight photometric bands sample broad spectral features that probe a wide range of molecular species (H$_{2}$, CO$_{2}$, CO, CH$_{4}$, HCN, CN, polycyclic aromatic hydrocarbons (PAHs), as well as broad ice and mineralogical features) that are not measureable from the ground due to telluric absorption and variability.  NIMBUS utilizes an optical design employing a tree of dichroics and fold mirrors to image multiple spectral band passes onto different regions of a single detector.  Each spectral band features a wide FOV to enable 1) decorrelation of time-variable effects using nearby reference stars, and 2) the ability to image extended sources.  Precise, real-time calibration of instrumental and atmospheric systematic variability using bright reference stars in the FOV enables high-precision spectrophotometric observations (S/N $>$ 10$^{4}$) of moderately bright sources (K$_{mag}$ $<$ 11) in one hour.  The instrument produces a three-dimensional (3D) data cube for each integration (two-dimensional --- 2D --- images in eight different wavelength bands), providing many of the efficiency benefits of a low-resolution integral field spectrograph.  Moreover, NIMBUS's design is optimized to take advantage of a large aperture in the stratosphere for measuring molecular bands in the NIR.  The optical design maximizes precision by eliminating many of the problems that preclude high-precision measurements with traditional slit and integral field spectrographs.  

\section{SCIENCE CAPABILITIES}
\label{sec:science}

NIMBUS provides access to molecular features unavailable to ground-based wide-field imagers due to atmospheric absorption (see Figure~\ref{fig:atm_trans}).  In addition, the NIMBUS instrument is designed to provide the advantages in photometric precision and spatial coverage of a wide-field imager while also providing the efficiency and internal calibration functionality of a low-resolution spectrograph.  The wide FOV and highly precise measurement capabilities for simultaneous multiple narrow-band filters are optimized for detecting molecular features that represent new science opportunities in a wide range of astrophysics and planetary science fields, including:

\begin{figure}
\begin{center}
\begin{tabular}{c}
\includegraphics[height=8cm]{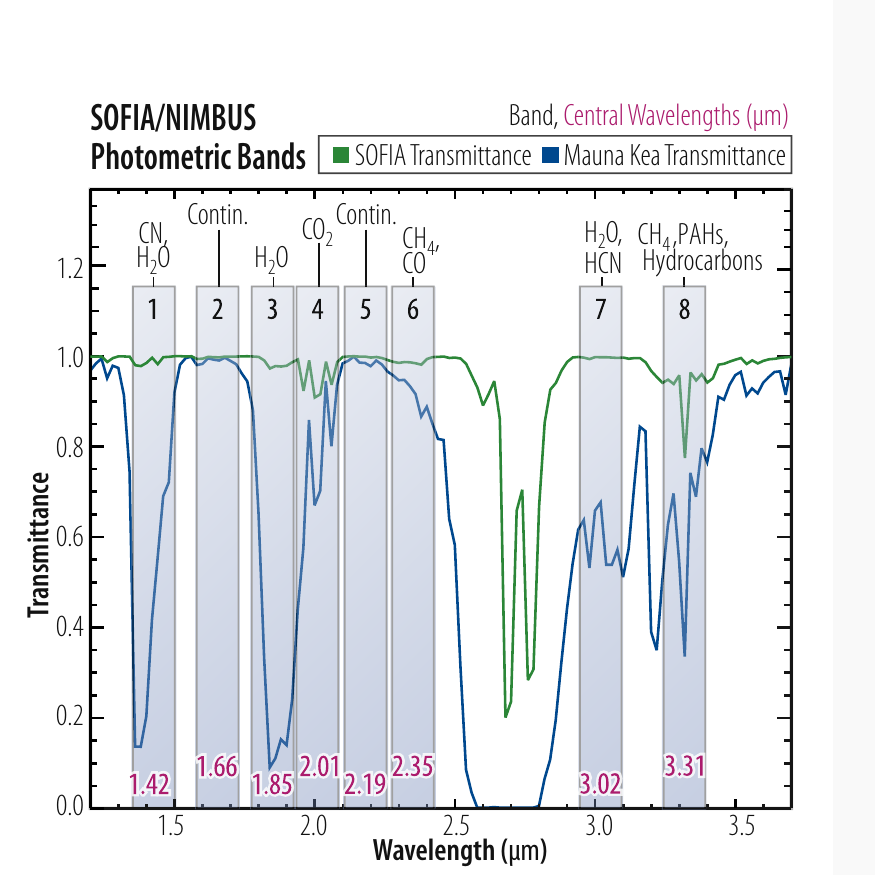}
\end{tabular}
\end{center}
\caption[example]{ \label{fig:atm_trans} The NIMBUS bands are positioned to target spectral features in specific regions of the NIR spectrum that are obscured from the ground by telluric absorption from H$_{2}$O, CH$_{4}$, and CO$_{2}$. With NIMBUS we can directly measure the intensity of these species (and many others) with a precision impossible from the ground. Each bandwidth is set to 0.16 $\mu$m in order to optimize the measurement of broad molecular features while avoiding overlap.}
\end{figure} 

\begin{itemize}

\item Characterization of the temperature structure, chemistry, and dynamics of extrasolar planet atmospheres during transit and occultation for a wide range of planetary and stellar masses

\item Measurements of spectral variability in ice and mineralogic spectral features for TNOs and Main-Belt asteroids indicative of hemispheric changes in surface composition

\item Spatially resolved maps of molecular constituents in cometary comae that would explore the dynamic release and distribution of gas for different comet classes

\item Identification and atmospheric characterization of new brown dwarfs in young stellar clusters and measurements of variability in nearby brown dwarfs    

\item Measurements of chemical gradients in globular clusters (GCs) and nearby galaxies to provide new insights into the evolution of stellar populations

\end{itemize}

The requirement to achieve a precision of 10$^{-4}$ for a multi-hour photometric time series drives a number of the critical design choices made in the design of NIMBUS. An important instrument design choice was to avoid photometric variability due to slit losses or slicing losses from a traditional spectrograph or an integral field spectrograph, which have been shown to produce minimum photometric errors on the order of 1 part in 10$^{3}$ in continuum regions\cite{knutson_et_al2007c}. The large SOFIA PSF combined with limited pointing stability from the telescope optical assembly magnifies spectrograph slit and slicing difficulties. A secondary aspect of the design is the ability to self-calibrate each exposure using one or more nearby reference stars.  This is especially important when attempting to measure molecular features that are also present in the telluric spectrum\cite{mandell_et_al2011}.  

Recent pioneering work on ground-based observations of exoplanet transits\cite{croll_et_al2010,croll_et_al2010b,croll_et_al2011,bean_et_al2010,bean_et_al2011} has demonstrated that simultaneous measurements of references stars and the use of an extremely wide aperature allow photometric precision on the order of several parts in 10$^{4}$ in regions without significant telluric contamination from the ground. However, these ground-based measurements encounter difficulties in regions with high telluric absorption, and these regions have errors increasing above 5$\times$10$^{-3}$ (J. Bean, priv. comm.). The Croll et al. results utilized a wide-field imager with custom-designed filters, while the Bean et al. results utilized several different near-infrared multi-object spectrographs.  

\subsection{Exoplanet Spectroscopy}

As planets transit their stars, NIMBUS's precise time-domain spectrophotometric observations can be used to probe the characteristics of the planetary atmospheres. This is accomplished by measuring the in-transit and out-of-transit stellar flux in order to detect small, wavelength-dependent changes as the exoplanet passes in front of or behind its parent star. Until the launch of the \textit{James Webb Space Telescope} (JWST), NIMBUS would be the premier instrument for measuring NIR molecular features in the spectra of exoplanet atmospheres.

The locations and width of the NIMBUS bands are optimized to measure key diagnostics for probing the structure and composition of exoplanet atmospheres. Water vapor is predicted to be a major component of nearly all transiting exoplanet atmospheres, and the NIMBUS bands are placed to measure warm and hot water vapor as well as other key molecular constituents (e.g., CH$_{4}$, CO$_{2}$, CO). By simultaneously observing multiple bands across a wide spectral range, NIMBUS would measure the composition and temperature structure of the atmospheres of exoplanets ranging from several times the mass of Jupiter down to a few times the mass of Earth. The NIMBUS bands simultaneously cover a spectral bandpass ($\Delta$$\lambda$/$\lambda$) of more than 100$\%$, compared to $\sim$20$\%$ for a single $J$, $H$, or $K$-spectral band. This broad spectral coverage and highly-efficient sampling capability paired with simultaneous photometric calibration would enable new insights in comparative planetology that would help prepare for more detailed measurements with JWST.

Even the hottest exoplanets emit most of their thermal flux in the infrared, and the strong molecular bands in planetary atmospheres lie longward of 1 $\mu$m; therefore, this is an important spectral range for exoplanet atmospheric characterization. Space-borne observatories present an advantage for long-wavelength observations due to reductions in thermal backgrounds and the absence of telluric absorption and variability. For these reasons, instruments such as NICMOS on the \textit{Hubble Space Telescope} (HST) and the IRAC camera on \textit{Spitzer} have been tremendous resources for exoplanet observations. However, due to its Earth-trailing orbit, \textit{Spitzer} will become unavailable for further observations in the near future (2013). NIMBUS would provide eight simultaneous spectral data points with similar precision as IRAC; however, IRAC on Warm \textit{Spitzer} can now only observe in a single spectral band at a time. HST is scheduled to remain on-line until $\sim$2020, but the current instrumentation limits HST to wavelengths shorter than 1.7 $\mu$m. Until the launch of JWST, NIMBUS would provide the only instrument capable of efficiently measuring the strong NIR molecular features in exoplanet atmospheres, in an environment where impact of telluric absorption is negligible.
NIMBUS would answer a number of outstanding exoplanet questions, including:

\begin{itemize}

\item \textbf{Characterization of Hot Jupiters}: Scientific mysteries surround these giant worlds, particularly their atmospheric composition\cite{fortney_et_al2006,swain_et_al2010,beaulieu_et_al2010,madhusudhan_et_al2011b}, atmospheric temperature structure\cite{fortney_et_al2006,burrows_et_al2007b,fortney_et_al2008,madhusudhan+seager2009,spiegel_et_al2009b,knutson_et_al2010}, and photochemistry\cite{zahnle_et_al2009}. The thermal and chemical structures of exoplanetary atmospheres combine to generate the emergent radiation; it is therefore essential to constrain both simultaneously. Currently, theoretical models require thermal inversions to fit the sparse observational data points for some hot Jupiters; however, a thermal inversion requires a strong absorber of visible wavelengths in the upper atmosphere, and the nature of any high-temperature optical absorber in hot exoplanets is still unknown. Although titanium oxide (TiO) has been suggested as the culprit\cite{hubeny_et_al2003,fortney_et_al2008}, it is not clear that vertical mixing in the radiative zones of hot Jupiter atmospheres is vigorous enough to keep a heavy molecule such as TiO aloft\cite{spiegel_et_al2009b}. Precise observations of the thermal radiation from these planets' atmospheres can firmly identify the planets with temperature inversions. Furthermore, comparison of the incident flux on a planet with the emergent day-side flux constrains the degree of heat transport to the nightside and thereby informs our understanding of the planet's atmospheric dynamics. NIMBUS measures the thermal and chemical structure of atmospheres by observing the emission from within molecular bands (which probe the upper atmosphere) as well as outside of molecular bands (which probe deeper into the atmosphere), and is therefore sensitive to both changes in temperature structure and changes in chemical composition (see Figure~\ref{fig:hotJup}). Assembling a wide-ranging census of the thermal and chemical structures of exoplanets would constrain which planetary characteristics lead to thermal inversions and aid in developing a chemical taxonomy of planetary atmospheres.

\begin{figure}
\begin{center}
\begin{tabular}{c}
\includegraphics[width=0.36\linewidth]{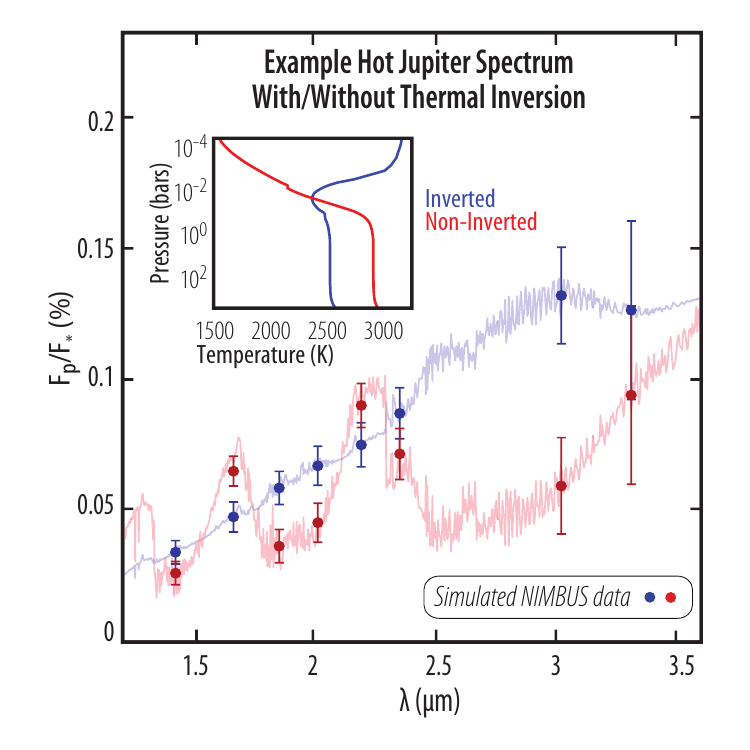}
\includegraphics[width=0.48\linewidth]{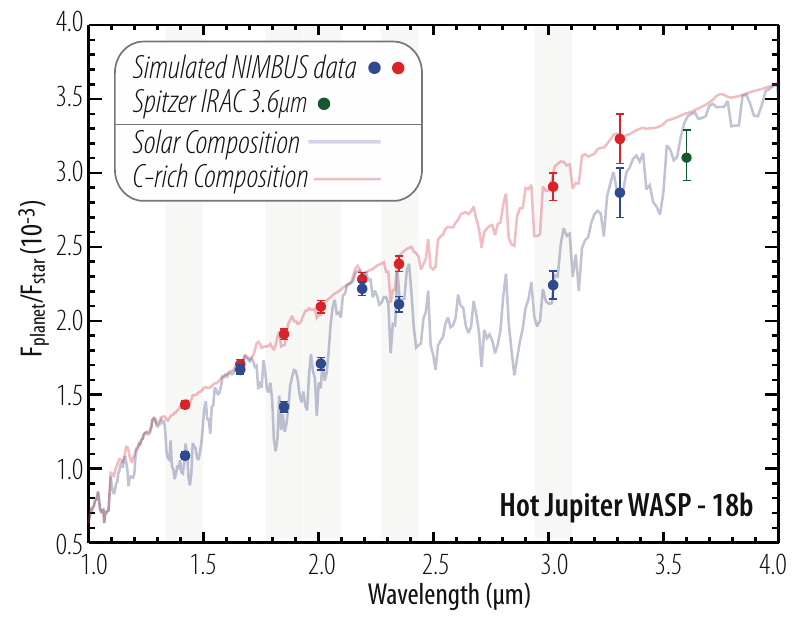}
\end{tabular}
\end{center}
\caption[example]{ \label{fig:hotJup}  NIMBUS would probe the regions of the NIR spectrum that are most sensitive to changes in the temperature and composition of exoplanet atmospheres. \textit{Left image}: NIMBUS would be able to identify thermal inversions in hot Jupiter atmospheres.  \textit{Right image}: Models of the ultra-hot Jupiter WASP-18b, with simulated NIMBUS data and errors overlayed. NIMBUS would be able to confirm a possible new class of carbon-rich planets, a discovery impossible from the ground (shaded gray bars).}
\end{figure}

\item \textbf{Characterization of Super-Earths}:  Recent studies have shown that super-Earth-mass planets may have a variety of compositional structures that are consistent with their measured masses and radii\cite{valencia_et_al2007,rogers_et_al2010,nettelmann_et_al2011}. However, since the bulk composition presumably influences the atmospheric chemistry, constraints on the atmospheric composition would help to break degeneracies in models of the interior structure. For the case of GJ 1214b, a 6.4 Earth-mass transiting planet orbiting an M-star, the mass and radius of the planet allow for a variety of interior compositions\cite{miller-ricci_et_al2009a,rogers_et_al2010}. The limited atmospheric observations available for this planet have yielded contrary conclusions from different groups ranging from a Neptune-like hydrogen-dominated atmosphere to one dominated by heavy molecules, similar to some terrestrial atmospheres in our own Solar System. The reduced molecular weight of an atmosphere containing molecular hydrogen inflates the geometric scale height and increases the amplitude of variation in transit radius with wavelength, while a heavier terrestrial-like (or water-rich) atmosphere leads to very little variation in the NIR spectral region\cite{miller-ricci_et_al2009a}. NIMBUS would resolve these degeneracies by providing high-precision NIR measurements in a number of critical bands at once (see Figure~\ref{fig:superEarth}).

\begin{figure}
\begin{center}
\begin{tabular}{c}
\includegraphics[width=0.65\linewidth]{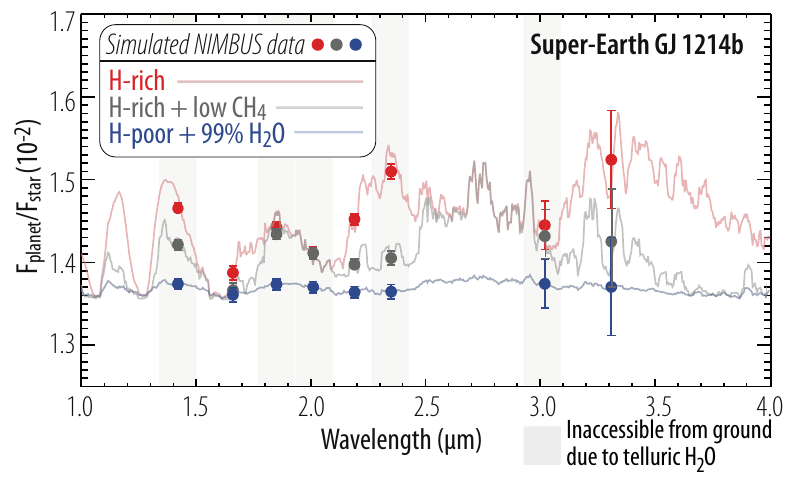}
\end{tabular}
\end{center}
\caption[example]{ \label{fig:superEarth} Models and simulated data for the super-Earth GJ 1214b with differing compositions. NIMBUS would be able to differentiate between H$_{2}$O-rich and H-rich atmospheres.}
\end{figure} 

\item \textbf{Constraints on Formation Scenarios}: Given high-confidence constraints on temperature structures and atmospheric abundances, it is possible to constrain the conditions under which a transiting planet formed. For example, the relative abundances of atomic species such as C, O, and N in a gas giant atmosphere can constrain the region of the protoplanetary disk in which the planet formed\cite{lodders2004,mousis_et_al2011,madhusudhan_et_al2011b}. In a protoplanetary disk, these heavy elements are incorporated into volatile molecules with varying evaporation temperatures, and they will therefore condense from a gas to a solid at different distances from the central star. By measuring the carbon and oxygen abundances (see Figure~\ref{fig:hotJup}) and bulk atmospheric density and composition of a number of planets, NIMBUS observations can constrain the commonality of formation location and mechanisms (core accretion vs. disk instability).

\item \textbf{Kepler Follow-Up and Pre-JWST Characterization}: NIMBUS would provide follow-up capabilities for the characterization of transiting planets identified by the \textit{Kepler Space Telescope} (Kepler). NIMBUS can characterize $\sim$40$\%$ of the first 19 confirmed Kepler planets, suggesting that the eventual sample of Kepler targets accessible with NIMBUS would be large. Planets on a wide range of orbits will provide new opportunities to constrain formation scenarios and the effects of stellar irradiation. The transit depths for planets on large orbits will be slightly smaller ($\sim$1$\%$ to a factor of a few) due to reduced atmospheric scale heights, but many of these cooler exoplanets would be well within the reach of NIMBUS, opening up a new class of transiting planets for characterization. Furthermore, SOFIA/NIMBUS would be able to fly to the ideal location to observe the transit and provide high-precision measurements in a single observation. This is particularly important if the transit only occurs once each year.   

\end{itemize}

\begin{figure}
\begin{center}
\begin{tabular}{c}
\includegraphics[width=0.65\linewidth]{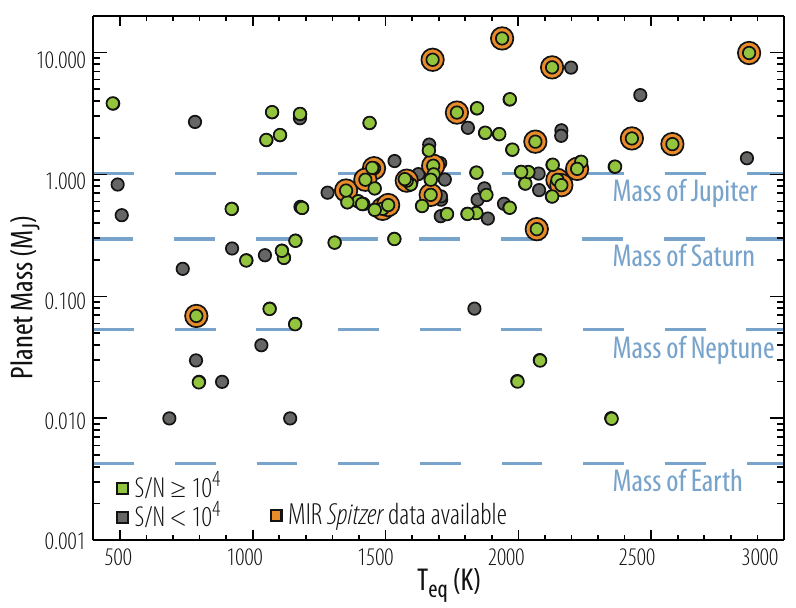}
\end{tabular}
\end{center}
\caption[example]{ \label{fig:currentExoplanets} NIMBUS characterizes planets around stars with sufficient flux to provide S/N $>$ 10000 (74 planets). NIMBUS can sample the full distribution of planet mass and equilibrium temperature (based on the incident stellar flux) for all known transiting planets. Complimentary Spitzer observations at longer wavelengths exist for 22 of the planets observable with NIMBUS.  New transiting exoplanets will increase the size of this sample.}
\end{figure}

There are currently 110 confirmed transiting exoplanets; NIMBUS has sensitivity to a planet-to-star flux ratio of $<$ 10$^{-4}$ for 74 currently known planets. Figure~\ref{fig:currentExoplanets} shows the masses and equilibrium temperatures for the known transiting planets, demonstrating NIMBUS would be sensitive to planets spanning the full range of masses and atmospheric temperatures. Furthermore, NIMBUS would provide high-precision observations of 22 systems that have complimentary longer wavelength observations with \textit{Spitzer}, providing a set of planets with high-precision photometry covering the NIMBUS bands as well as \textit{Spitzer} bands between 3.6 and 24 $\mu$m.  The characterizable exoplanet catalogue will continue to grow as ground-based wide-field surveys for nearby short-period exoplanets--including the highly-successful Hungarian Automated Telescope Network (HATNet and HAT South), Super Wide Angle Search for Planets (SuperWASP), and recent MEarth and HAT-South surveysÑmake new discoveries. This catalogue would be expanded further if a space-based all-sky survey for nearby transiting planets, such as the Transiting Exoplanet Survey Satellite (TESS; currently in Phase A development for the Explorer Program), were approved.

\subsection{Sensitivity}

\begin{figure}
\begin{center}
\begin{tabular}{c}
\includegraphics[height=10cm]{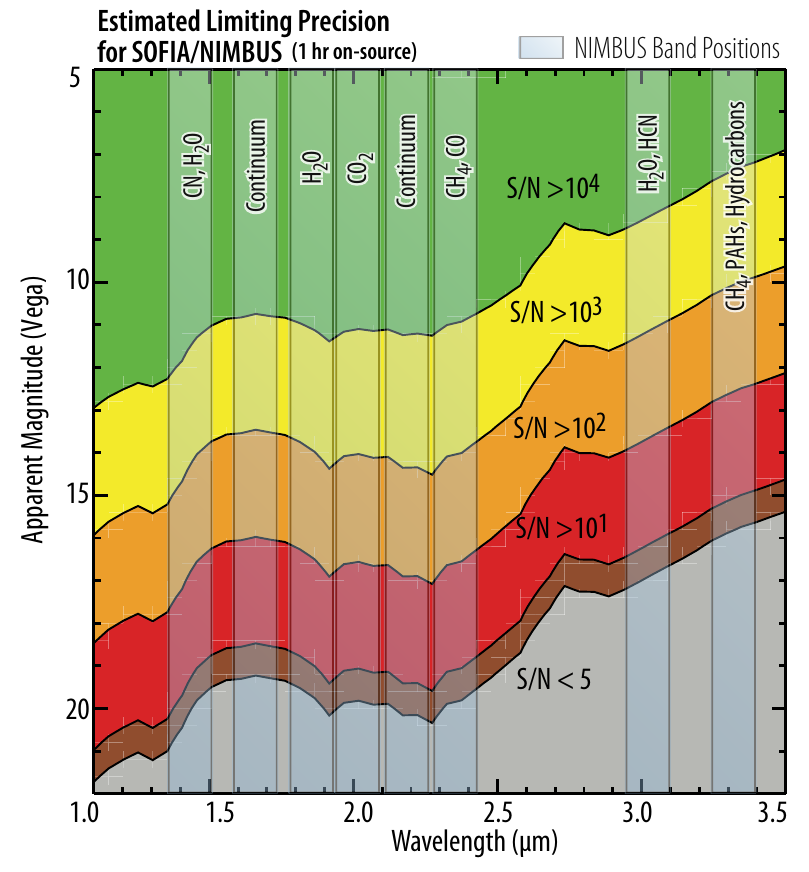}
\end{tabular}
\end{center}
\caption[example] 
{ \label{fig:sensitivity} NIMBUS sensitivity with wavelength and source brightness was estimated with an instrument simulator. The positions for the eight NIMBUS bands are marked with vertical blue bars. For the six short-wavelength bands, NIMBUS has a limiting magnitude of K$_{mag}$ = 20 in one hour, and NIMBUS can reach S/N $>$ 10$^{4}$ for K$_{mag}$ $<$ 11 in one hour.}
\end{figure}

In order to estimate the relative contributions from the possible noise sources inherent in various instrument designs, the NIMBUS team developed an end-to-end instrument simulator incorporating a suite of parameters based on the SOFIA observatory and the instrument design constraints. The simulator calculates the photon flux through the SOFIA aperture, beginning with a Kurucz spectral model (of the desired spectral type) that is scaled to the desired apparent magnitude.  The simulator then incorporates a model of the various contributions of the atmospheric emission and absorption at SOFIA flight altitudes, the telescope optical assembly, and the characteristics of the instrument. A line-by-line radiative transfer model (LBLRTM)\cite{clough_et_al2005} and an atlas of OH emission lines\cite{rousselot_et_al2000} were used to model the transmission and radiance of the atmosphere. The model produces high-resolution transmittance and radiance spectra at a specified altitude, and these spectral points are integrated to represent the specific bandpass regions used by NIMBUS. The telescope model includes the mirror size, emissivity, the FWHM of the PSF at different NIR wavelengths, and the image motion across different pixels due to pointing inaccuracies. The instrument model includes the optical throughput and the quantum efficiency of the detectors, the pixel size and required photometric aperture size for precision measurements (2$\times$ FWHM), and the detector read-noise, depth of the full well, and minimum exposure time.  The simulation algorithm produces a S/N for a specific object at any bandpass between 1 and 5 $\mu$m, and by incorporating the noise contribution from additional calibration sources, the team accurately estimated the precision for an actual observation. A plot of NIMBUS's sensitivity is shown in Figure~\ref{fig:sensitivity}. The photometric error budget is shown in Table~1, demonstrating how the 10$^{-4}$ precision stability is achieved for a G2V star with K$_{mag}$= 11 in one hour of observing time. The resulting photometric precision would be limited by the fainter of the science target and the calibration star. For the brightest targets, NIMBUS could achieve 4$\times$10$^{-5}$ precision in one hour.

The instrument simulator was used to evaluate a wide range of NIR instrument designs, including simultaneous multi-band imaging with dichroics, traditional spectroscopy, and integral field spectroscopy techniques. Beyond the stochastic noise (dominant at short wavelengths) and the telescope and sky background at longer wavelengths, the largest potential error source is the jitter of the image across multiple pixels (contributing uncertainties of 1 part in 10$^{5}$ using a wide aperture). Using a traditional spectrograph with a slit width equivalent to the expected SOFIA FWHM and the most optimistic pointing stability, the variable slit losses would lead to an uncertainty of 0.5\% in the final photometric time series; though these errors could be corrected somewhat through PSF modeling and decorrelation techniques, the final photometric stability would be insufficient to meet the science goals.  A slitless spectrograph similar to the one baselined for WFIRST was closely considered since it would meet the baseline design goals (no slit losses and a wide FOV), but the thermal backgrounds from adjacent spatial locations on the sky limit the sensitivity for each of the spectral channels. The dichroic design enables the instrument to simultaneously image in several bands while maintaining a very high instrument throughput ($\sim$40\%, including the detector efficiency).

\begin{table}[h]
\label{tab:ErrorBudget}
\begin{center}       
\begin{tabular}{lcccc} 
\hline
\rule[-1ex]{0pt}{3.5ex}  Item & Instrument Requirements & & Level of Precision  \\
\rule[-1ex]{0pt}{3.5ex}   &  & Band 1 & Band 6 & Band 8  \\
\rule[-1ex]{0pt}{3.5ex}   &  & (1.42 $\mu$m) & (2.35 $\mu$m) & (3.31 $\mu$m)  \\
\hline
\hline
\rule[-1ex]{0pt}{3.5ex}   Photon Poisson Noise & Primary Mirror 2.5 m & 4$\times$10$^{-5}$ & 4$\times$10$^{-5}$ & 1$\times$10$^{-4}$  \\
\hline
\rule[-1ex]{0pt}{3.5ex}   Atmospheric Scintillation &  & 3$\times$10$^{-5}$ & 3$\times$10$^{-5}$ & 3$\times$10$^{-5}$  \\
\hline
\rule[-1ex]{0pt}{3.5ex}   Read Noise & SUTR read-out & 9$\times$10$^{-6}$ & 2$\times$10$^{-5}$ & 6$\times$10$^{-5}$  \\
\hline
\rule[-1ex]{0pt}{3.5ex}   Telescope Thermal & $\epsilon$$\sim$0.1; T $<$ 245 K & 2$\times$10$^{-8}$ & 5$\times$10$^{-5}$ & 2$\times$10$^{-3}$  \\
\hline
\rule[-1ex]{0pt}{3.5ex}   Sky Radiance Background & & 5$\times$10$^{-5}$ & 5$\times$10$^{-6}$ & 5$\times$10$^{-4}$  \\
\rule[-1ex]{0pt}{3.5ex}   Detector Stability & QE \& pix-to-pix variations & 1$\times$10$^{-5}$ & 2$\times$10$^{-5}$ & 2$\times$10$^{-5}$  \\
\hline
\hline
\rule[-1ex]{0pt}{3.5ex}   Total & & 7$\times$10$^{-5}$ & 9$\times$10$^{-5}$ & 3$\times$10$^{-3}$  \\
\hline
\end{tabular}
\caption{NIMBUS error budget for a K$_{mag}$=11, G2V star in one hour of observing time achieves 10$^{-4}$ precision, while brighter stars would achieve even better photometric precisions.} 
\end{center}
\end{table} 

\section{INSTRUMENT DESIGN}
\label{sec:instrument}

The key requirements for the NIMBUS instrument design are derived from the most stringent requirements associated with the science goals, which mandate high spectrophotometric sensitivity in the NIR, multiple simultaneous photometric bands, and a field of view sufficient to provide at least one bright reference star. To meet these requirements NIMBUS is implemented as a multicolor imager employing dichroics to separate the spectral bands. This approach yields a simple and easily implemented design that would be able to meet the challenge of NIR spectrophotometric characterization in the SOFIA operational environment over a variety of flight plans and operating altitudes. NIMBUS therefore has no strict altitude or water vapor depth requirements beyond the requirement of observations at stratospheric altitudes ($>$ 33,000 ft). 

NIMBUS interfaces directly to the standard SOFIA optical assembly to receive the telescope beam.  A K-mirror before the instrument keeps a fixed position angle on the detector for all observations of a given target.  The instrument fore-optics place a field mirror at the telescope focus and include a set of relay optics used to reduce the instrument size. The field-corrected beam is collimated and immediately split into spectral bands using a tree of dichroics. The selection of central wavelengths and spectral bandwidths are chosen to maximize the capability for differential measurements between regions of the spectrum with and without major molecular bands. The NIMBUS bands are selected to cover gas-phase rovibrational bands of H$_{2}$O, CH$_{4}$, CO$_{2}$, CO, HCN, and CN, as well as several continuum regions. The bandwidth is the same for each of the bands, and is chosen to provide the most contrast between the bands while avoiding overlap in regions with closely-space features. The dichroic tree separates images into short and long wavelength channels, where the short wavelength channel (SWC) image onto a 2.5 $\mu$m cutoff Hawaii 2RG detector, while the long wavelength channel (LWC) images onto a 5.3 $\mu$m cutoff Hawaii 2RG detector. Each detector has four bands arranged in a 2$\times$2 matrix.

NIMBUS FOV is driven by the minimum field size that would allow at least one simultaneous reference star bright enough to provide real time independent monitoring of photometric stability during NIMBUS observations. Characterization of exoplanet atmospheres is a primary science goal for NIMBUS and requires precision spectrophotometry at the level of 10$^{-4}$ in 1 hour of time on-sky in the $K$-band for objects brighter than 11$^{th}$ magnitude. A 6 arcmin $\times$ 6 arcmin FOV and 8 total photometric bands yields the optimum trade for a reference star with K$_{mag}$ $<$ 11, as confirmed by the space density of calibration sources as a function of galactic latitude in the 2MASS point source catalog.

Wide-field simultaneous imaging in multiple bands requires large format detectors. NIMBUS leverages the large investment made in the development of IR arrays for JWST. The Teledyne Hawaii 2RG (H2RG) has 2048$\times$2048 pixels with a pixel pitch of 18 $\mu$m and is available to the general astronomical IR community. To ensure an oversampled image, the team adopted a 0.4 arcsec pixel scale, given the minimum 80$\%$ encircled energy SOFIA PSF size of 2.9 arcsec at 3.0 $\mu$m. Further oversampling is available with the NIMBUS design if required by simply defocusing the SOFIA secondary mirror; NIMBUS is therefore largely insensitive to the uncertainty in the actual SOFIA PSF. The pixel scale and FOV requires $\sim$900 pixels for each of the spectral images, which allows four images to be simultaneously imaged on the same Hawaii 2RG detector. 

The NIMBUS instrument design is straightforward with only a single mechanism for a field derotator (i.e., K-mirror). Dichroics split the beam into eight separate wavelength bands, which are then imaged by two Teledyne H2RG detectors. The optics system and detector assemblies are packaged into a custom dewar. The interior of the dewar uses G10 struts that thermally isolate the optical support structure from the ambient-temperature dewar outer shell. A closed-cycle cryocooler could remove heat from the internal dewar structure and maintain operational temperatures of 78$\pm$5 K for the optical support structure and 38.00$\pm$0.005 K for the detectors. The entire instrument would be controlled by commercial electronics with custom software for interfacing with SOFIA and users.

\subsection{Optical Design}
\label{subsec:optical}

The NIMBUS science objectives drive the optical design specification to simultaneously obtain eight images of the field being observed with with scientifically-selected wavelength bands.  The field requirement is derived from the need to find a bright reference star near the target star.  Within a 6 arcmin diameter, NIMBUS would find a reference star of $K$$<$11 to obtain the specified photometric accuracy of 10$^{-4}$ in 1 hour of exposure.  The PSF of the optical system varies by $<$ 2 arcsec over the operational wavelength range, while field pointing is maintained by the SOFIA telescope's pointing stability.  The optical requirements are that NIMBUS accepts the SOFIA beam (F/\#=19.66 reduced to F/\#=3.4 with demagnification), produces eight spectral bands, operates from 1.34-3.39 $\mu$m, maintains image quality better than 2 arcsec at all wavelengths, and uses H2RG 2k$\times$2k detectors with 18 $\mu$m pixels.

The NIMBUS optical system uses dichroic beam splitters to separate eight simultaneous wavelength bands. This maximizes the system's optical throughput while obtaining eight spectral images over a very large spectral band pass. The dichroics achieve a high throughput by directing the photons in each wavelength band to its image only. Four images in each channel share one detector and one set of camera optics. The eight bands are split into two channels (based on wavelength), which produces a short-wavelength channel (1.34--2.09 $\mu$m) and long-wavelength channel (2.11--3.39 $\mu$m). This spectral separation enables separate exposure time control because of the differences in the background levels of the two channels, and detector sub-arrays further improve this flexibility.

\begin{figure}
\begin{center}
\begin{tabular}{c}
\includegraphics[width=0.90\linewidth]{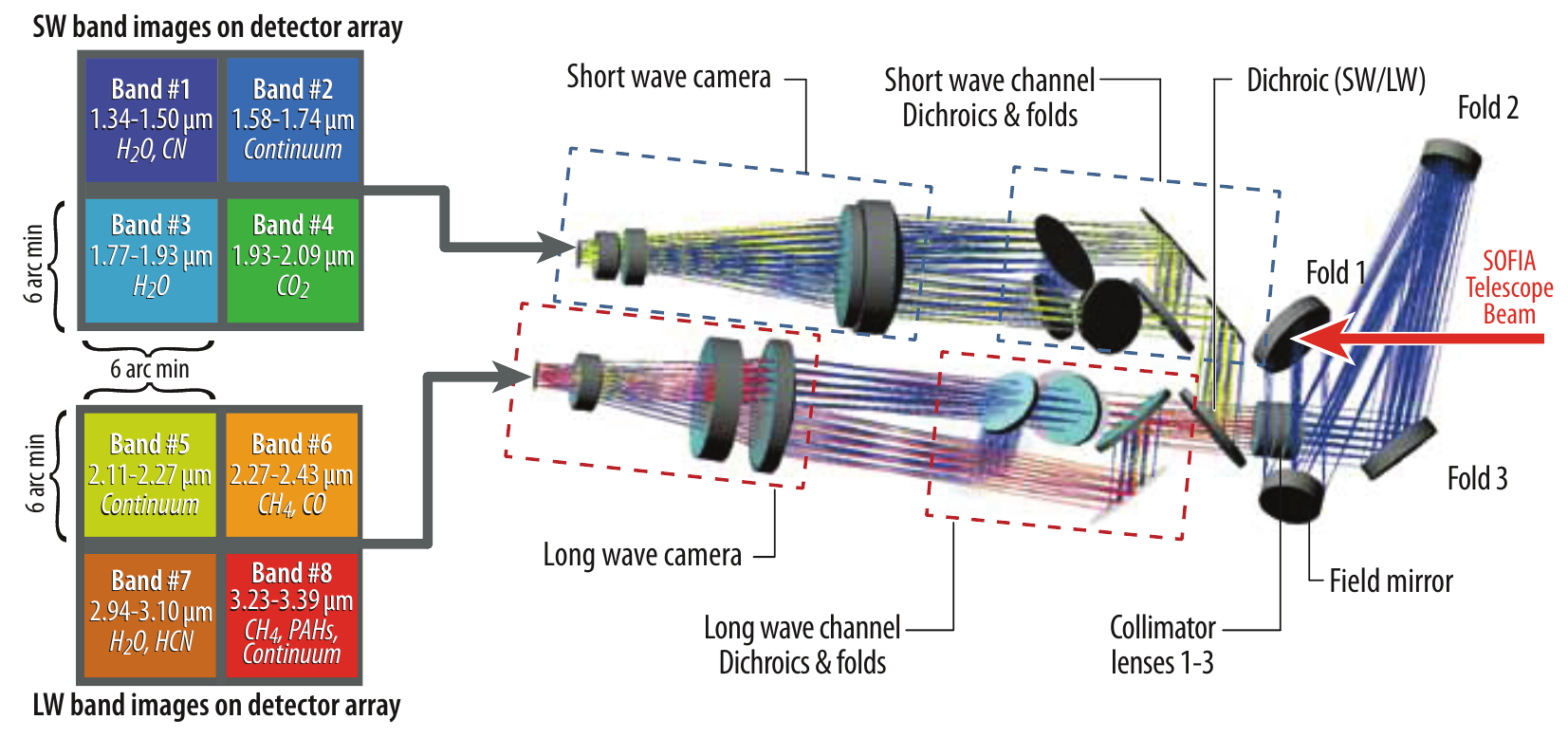}
\end{tabular}
\end{center}
\caption[example] 
{ \label{fig:optics} The NIMBUS optical layout has only one moving part making it easy to implement and calibrate.  The image rotator (not shown here) is used to maintain a fixed position angle on the detector.  The input beam is collimated and then dichroic beam splitters are used to separate eight simultaneous wavelength bands.  Four beams are passed through one set of camera optics for both the short-wave (SW) and long-wave (LW) arms of the instrument.  This design maintains high throughput over a wide field of view.}
\end{figure}

The baseline NIMBUS optical layout is shown in Figure~\ref{fig:optics}. The incoming beam from the SOFIA telescope is incident on a K-mirror that maintains a constant position angle on the detector.  Then a fold mirror reflects the beam to a field mirror located at the telescope focus. Two additional fold mirrors relay the beam to a three-lens collimating assembly. The beam is split after the collimators so that no image quality degredation is introduced by the dichroics, and the image position is insensitive to the optical path length in collimated space. After collimation, a dichroic splits the beam into short and long wavelength channels. Then the short and long wavelength beams are separated into four wavelength bands in the collimated space. The camera in each channel images the four separate collimated beams onto a single H2RG science camera assembly. The image position on the detector array is determined by the beam angle in the collimated space, which is accomplished by the position of the three fold mirrors in front of each camera. All optical components use IR materials with sizes and shapes readily obtainable from several commercial vendors. 

\begin{figure}
\begin{center}
\begin{tabular}{c}
\includegraphics[width=0.6\linewidth]{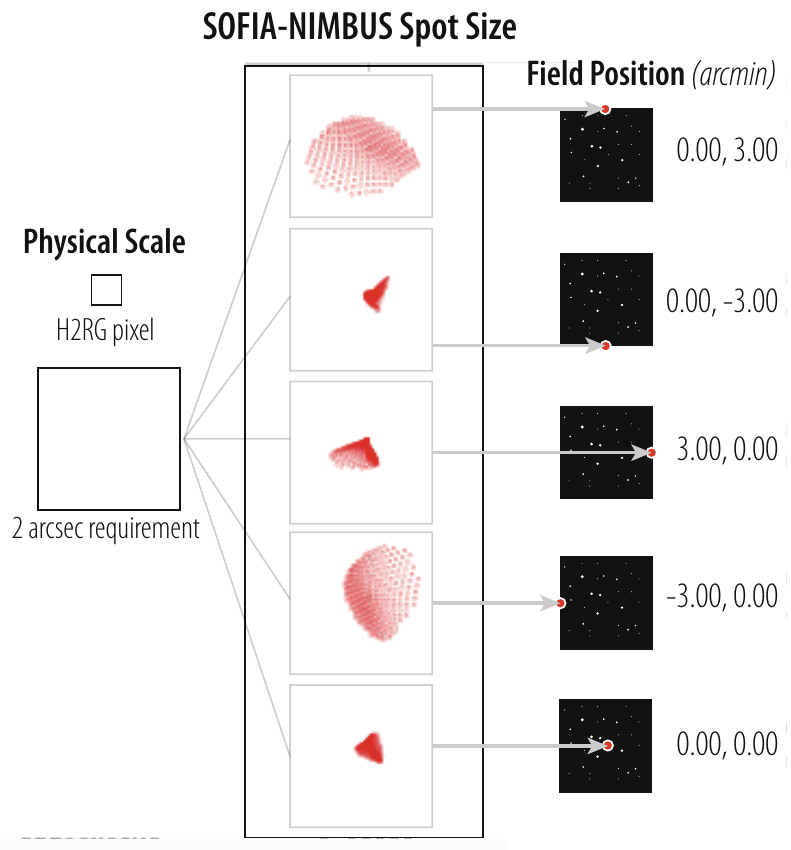}
\end{tabular}
\end{center}
\caption[example] 
{ \label{fig:spots}  NIMBUS spot diagrams verify the optical design meets image quality requirements.}
\end{figure}

The image quality (spot size no more than 2 arcsec RMS) is an important requirement for meeting the science objectives. With a NIMBUS plate scale of 0.4 arcsec, 2 arcsec are equivalent to 90 $\mu$m on the detectors. The NIMBUS PSF over the whole FOV and at all wavelengths meets the requirements of 90 $\mu$m RMS. Figure~\ref{fig:spots} shows the spot diagrams at several field points within a typical spectral image, demonstrating that the spot sizes are within the NIMBUS instrument requirements.

NIMBUS optical system alignment tolerances are achieved by aligning the subassemblies as the instrument is integrated. The optical components that require alignment are between the focal plane arrays and the short wavelength and long wavelength camera lenses, and between the field mirror and the collimator. All components require 30 arcsec alignment tolerances to meet the required performance. For the camera optics, a F/3.4 beam in front of the detector produces a depth of focus of $\pm$50 $\mu$m. The focus is measured using an H2RG in the focal plane, and shims would be used to optimize the detector position with respect to the camera lens assembly. This process requires several iterations, including cool-downs of the test dewar, to achieve optimal focus. The field mirror and collimator would achieve cold alignment and focus using two Davidson D-275 auto-collimating alignment telescopes, two theodolites, and a cryogenic test dewar.

\subsection{Mechnical Design}
\label{subsec:mech}

\begin{figure}
\begin{center}
\begin{tabular}{c}
\includegraphics[width=0.45\linewidth]{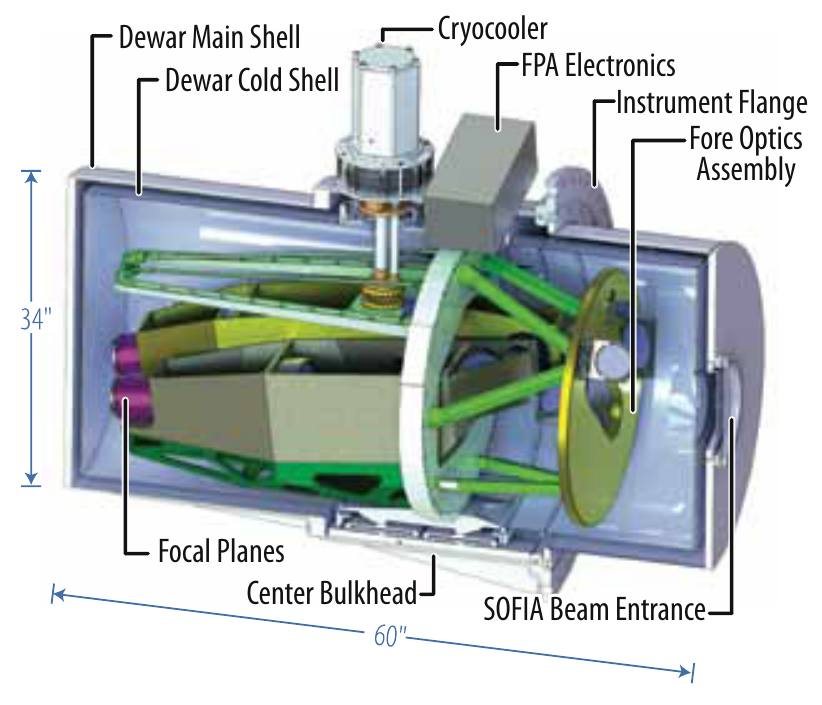}
\includegraphics[width=0.50\linewidth]{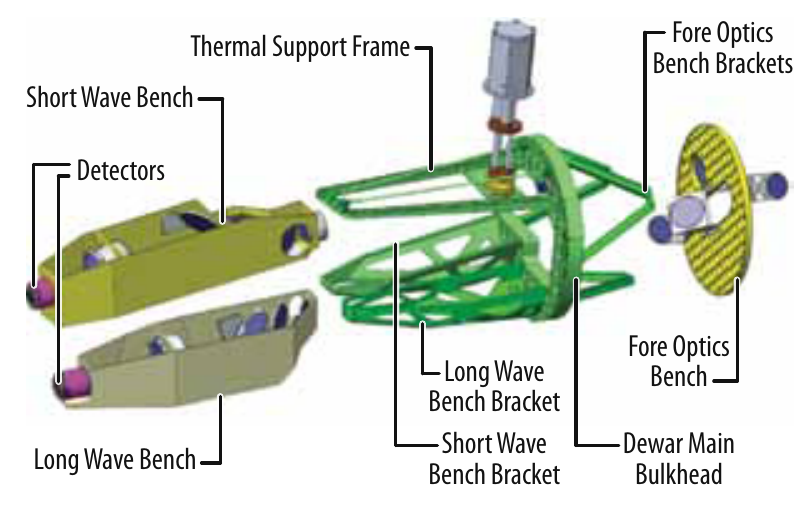}
\end{tabular}
\end{center}
\caption[example] { \label{fig:mech} \textit{Left image}: A cut-through of the dewar shows the instrument layout. \textit{Right image}: Exploded view of the NIMBUS instrument, showing the internal structure, three optical benches, and instrument sub-assemblies.}
\end{figure}

The NIMBUS mechanical design provides rigid mounting of the optics while maintaining a light weight so as to fit within SOFIA instrument mass constraints.  Figure~\ref{fig:mech} shows an internal view of the dewar components and an exploded view of the instrument sub-assemblies. The fore optics bench houses the field mirror and several folds. The LW and SW benches house the dichroic beam splitter and mirror trees, collimator and camera lens assemblies, baffles, and the focal plane arrays. Aluminum brackets support the optical benches. The benches are 6061T651 aluminum to minimize thermal expansion differences with the cold structure and optical mounts and to maximize thermal conduction.

The cold structure supports the optical benches, cryocooler cold head interface, and electrical/thermal harnesses. A circular main bulkhead serves as the primary mechanical interface. The bulkhead is coupled to the dewar main shell structure by re-entrant G10 standoffs, providing an isolating thermal path to meet the cryogenic temperature requirements (See Sec.~\ref{subsec:thermal}). Machined 6061T651 aluminum brackets on both sides of the plate support the optical benches. An A-frame to support the thermal straps extends from the main bulkhead aft to the cryocooler second stage cold head and focal plane assemblies.

The NIMBUS optical mounts follow conventional cryogenic mounting designs.  The 80 mm diameter collimator lens group is barrel-mounted with spring-loaded retainers providing axial preload. Component centration is maintained with spacers. The 100-140 mm diameter beam splitters are clamped into their mounts with three-point spring-loaded retainers and centered with spacers. The mirrors are mounted the same way as the beam splitters. The larger 190 mm diameter camera lens groups are housed in aluminum lens barrels and clamped in place with spring-loaded retainers.

The H2RG focal plane array (FPA) mounts are attached directly to the lens barrels. A short tangent blade flexure interface connects the FPA camera housings to the lens barrels. The FPA harness is supported by nearby brackets to reduce harness loads on the FPA mount.

The NIMBUS dewar is based on the SOFIA/SAFIRE dewar design.  The NIMBUS dewar uses a simplified SAFIRE support structure and has only two shells--a main outer vacuum shell at ambient temperature and an internal cold shell.  The Focal Plane Electronics (FPE), are contained in a Leach ARC-170 box mounted to the main center shell of the dewar.

\subsection{Cryogenic \& Thermal Design}
\label{subsec:thermal}

The NIMBUS design uses off-the-shelf components and standard cryogenic practices to achieve operational temperatures. During normal operation, the cryogenics system would cool the optical bench below the required 78 K and the detectors to 38 K.  The NIMBUS dewar was designed to meet SOFIA interface requirements. This vacuum vessel meets FAA pressure vessel requirements and acts as a containment vessel in the event of a 9 g shock.  An internal radiation shield is mechanically supported by G10 standoffs that also provide thermal isolation from the $\sim$300 K vacuum vessel. This radiation shield is blanketed on the outside with Multi-Layer Insulation to minimize radiation from the warm vacuum vessel. The radiation shield also mechanically supports the internal optics and includes a number of features to minimize stray light: the interior is painted black, mechanical interfaces employ tongue and groove seals, and tortuous path pump out ports are utilized. The radiation shield, and all internal optics coupled to it, are actively cooled by the first stage of a two-stage cryocooler. The second stage is connected to the detectors via a high-conductivity heat strap.

\begin{figure}
\begin{center}
\begin{tabular}{c}
\includegraphics[height=8cm]{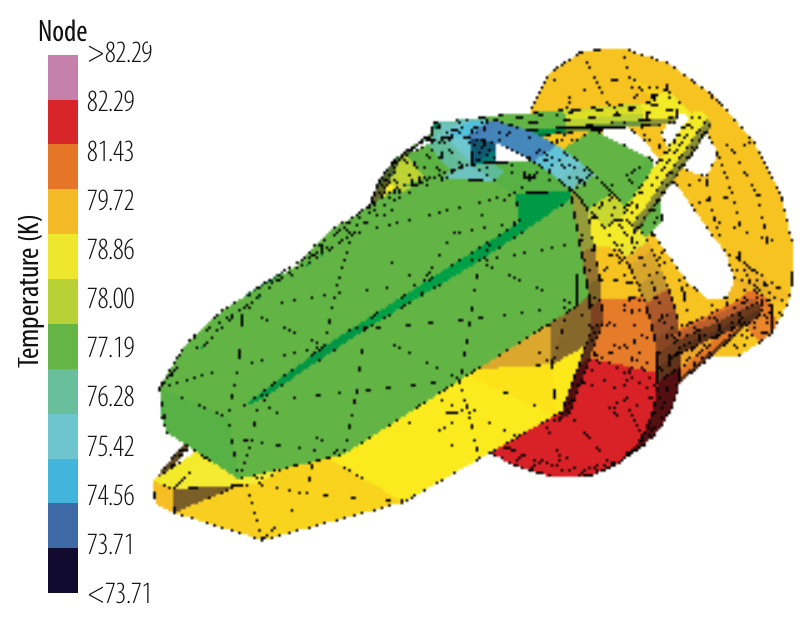}
\end{tabular}
\end{center}
\caption[example] 
{ \label{fig:thermal}  The thermal model demonstrates 78$\pm$2 K spatial gradients, which keeps the PSF positioned in the image plane.}
\end{figure}

The NIMBUS baseline cryocooler is the Cryotech PT810, which delivers adequate cooling power margin for a low cost. Other mechanical coolers or liquid cryogens could also be used to meet top-level NIMBUS science requirements. The Cryotech PT810 cryocooler is a COTS two stage pulse tube cooler, simultaneously providing $\sim$70 W of cooling at 78 K at the first stage and 25 W of cooling at 38 K at the second stage. Thermal analysis estimates $\sim$42 W of power (11 W through the G10 isolators, 31 W through the blankets) removed by the first stage while maintaining the optical bench at 78 K, with a maximum spatial gradient of about $\pm$2 K over the mechanical structure (Figure~\ref{fig:thermal}). The detectors dissipate heat on the order of mW and are further thermally isolated from the optics and internal structure; therefore, the second stage has more than sufficient capacity to cool the detectors, even with the long strap lengths from the detectors to the cryocooler.

The optical bench temperature is not actively controlled. The detectors are cold biased, using a PID-controlled heater to maintain the required 5 mK stability. While the pulse tube cooler has no mechanical moving parts at low temperature, the cold finger does have a displacement of 1 $\mu$m at the ÒbreathingÓ frequency ($\sim$1 Hz) of the high and low pressure He gas. Flexible copper foil straps are used at both stages to minimize vibration from the cold finger to the dewar internals. The He compressor draws 8 kVA of electrical power and requires chilled water for waste heat removal, which may be accomplished using a separate laboratory chiller. 

\section{CONCLUSIONS}

NIMBUS would be an excellent addition to the suite of instrumentation on the SOFIA airborne observatory.  This instrument would achieve 10$^{-4}$ photometric precision on K$<$11 stars in 1 hour, enabling new science programs over a broad range of astrophysics.  This instrument would be uniquely capable to characterize the atmospheres of transiting exoplanets in the era leading up to JWST.  It would also enable high precision photometric observations important for understanding TNOs, comets, brown dwarfs, and globular clusters.  The NIMBUS design uses traditional optical components in a new way to achieve ultraprecise photometric precision with high throughput.

\acknowledgments     
We would like to thank Klaus Hodapp, Ian McLean, Harvey Moseley, William Oegerle, Karl Stapelfeldt, and Motohide Tamura for stimulating and insightful discussions.  The authors would like to acknowledge the internal support provided by the Goddard Space Flight Center.  


\bibliography{biblio.bib}

\end{document}